\documentclass[conference,10pt]{IEEEtran}
\usepackage{amsfonts}
\usepackage{amssymb}
\usepackage{amsmath}
\usepackage{graphicx}
\usepackage{subfigure}
\usepackage{cite}\usepackage{setspace}
\usepackage{algorithm}
\usepackage{algorithmic}
\usepackage{mathrsfs}
\usepackage{subeqnarray}
\usepackage{cases}
\usepackage{color}
\usepackage{xcolor}
\usepackage{bbm}
\usepackage{bm}
\usepackage{flushend}
\begin{document}

\title{Joint Resource Block-Power Allocation for NOMA-Enabled Fog Radio Access Networks}

\author{\IEEEauthorblockN{Binghong Liu, and Mugen Peng~\IEEEmembership{Senior~Member,~IEEE} \\
State Key Laboratory of Networking and Switching Technology,\\
Beijing University of Posts and Telecommunications, Beijing, 100876, China. \\ Email: 2013210720@bupt.edu.cn\\
}}
\maketitle

\begin{abstract}
In order to achieve efficient communication in the fifth generation (5G) networks, non-orthogonal multiple access (NOMA) technique has been utilized in fog radio access networks (F-RANs). In this paper, we investigate the resource allocation problem in a NOMA-enabled downlink F-RAN. To maximize the weighted sum rate of NOMA users served by  fog-computing-based access points (F-APs), the resource block (RB) allocation and power allocation are optimized. Specifically, we decouple the problem into RB allocation and power allocation problems. The former is modeled as a many-to-one matching game and we propose a modified swap-enabled matching algorithm to solve it, which takes interference threshold into consideration. The later is a non-convex problem, we transform it into a tractable one via some approximations and get the closed-form expressions of power allocation coefficients. Finally, we combine the both to propose a joint resource allocation algorithm, which is preformed iteratively to obtain the optimal result. Simulation results are provided to show the performance of the algorithm.
\end{abstract}
\vspace{-0.25em}
\section{Introduction}
To support an ever increasing number of mobile devices and meet the drastically growth of mobile traffic in next generation wireless communication networks, fog radio access networks (F-RANs) [1][2] have been proposed as a promising architecture to provide high spectral efficiency as well as energy efficiency. To be more specific, the capabilities of local signal processing, cooperative radio resource management and distributed caching in edge devices have been fully exploited, which means that some user equipments (UEs) no longer access to the baseband unit (BBU) pool through fronthaul links and as a result, decreasing the traffic load on fronthaul links and relieve the heavy burden of large-scale signal processing on the centralized BBU pool.

NOMA is a spectrum-efficient technology for 5G communication networks [3]. It allows multiple users to simultaneously utilize the same time resource as well as frequency resource in a non-orthogonal way, which differs from the traditional orthogonal multiple access (OMA). At the transmitter side, it superposes all the signals requested by its serving users and transmit the superposed signal to users. While at the receiver side, in order to deal with the severe interference introduced at the transmitter side, successive interference cancelation (SIC) is applied to decode then subtract signals for other users and finally decode the own signal according to users' channel conditions.

Many researchers have focused on resource allocation in heterogeneous networks (HetNets)[4]-[6]. For instance, the authors in [4] investigate spectrum and power allocation problem between RRHs and high-power node (HPN) in a heterogeneous cloud radio access networks (H-CRANs) to maximize the energy efficiency of the whole system. In [5], authors model the power allocation problem in HetNets as a Stackelberg game and optimize the throughput of small cell BS and macrocell BS respectively while satisfying the constraints of user quality-of-service (QoS) requirements. In [6], cluster formation and resource allocation problems for NOMA-enhanced HetNets is studied. The authors take the power disparity and delay tolerance into consideration, which represents an imperfect NOMA scenario. Bi-partite matching and sequential convex programming techniques are exploited to solve the problem.

In this paper, we consider a NOMA-enabled downlink F-RAN and optimize resource allocation scheme of F-AP users (FUEs) to maximize the weighted sum rate of FUEs. Contributions of this paper can be summarized as follows:
\begin{itemize}
\item We formulate a resource allocation problem in NOMA-enabled downlink F-RAN, where NOMA technique is used by F-APs to enhance the spectral efficiency. We aim to maximize the weighted sum rate of FUEs via reasonable resource allocation schemes.

\end{itemize}
\begin{itemize}
\item We decouple the original problem into a RB allocation problem and a power allocation problem. The RB allocation problem can be modeled as a many-to-one matching game and solved by a modified swap-enabled matching algorithm. The power allocation problem is rewritten as a difference of convex function, which can be further transformed into a convex one via some approximations and we get the closed-form expressions of power allocation coefficients.
\end{itemize}
\begin{itemize}
\item Simulation results are provided to show the performance of the proposed algorithms, which also demonstrates that the performance of NOMA-enabled F-RAN is much more better than the conventional OMA-based F-RAN.
\end{itemize}

\vspace{0.5em}
\begin{figure}[htbp]
\centering
\includegraphics[width=0.45\textwidth]{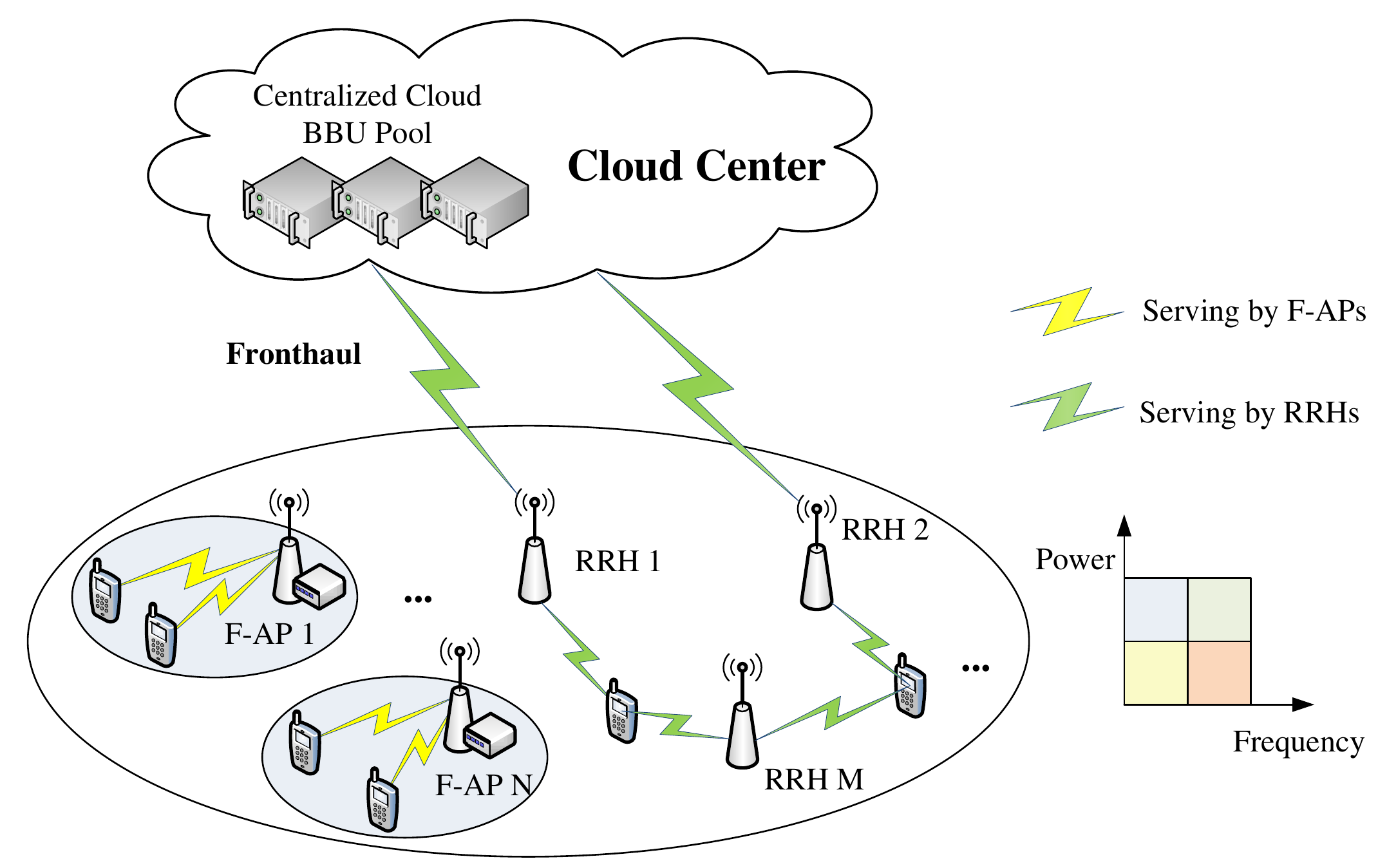}
\caption{NOMA-enabled Downlink F-RAN with one F-AP, $M$ RRHs.}
\end{figure}
\vspace{-0.9em}
\section{System Model}
Considering a NOMA-based downlink F-RAN system with $N$ F-APs, $M$ RRHs and multiple users. The set of F-APs is ${\rm{{\cal F}{\cal A}{\cal P}}} = \left\{ {1,2,...,N} \right\}$. The set of users served by RRHs is represented as ${\rm{{\cal R}{\cal U}{\cal E}}} = \left\{ {1,2,...,M} \right\}$. The bandwidth of the system is $B$ and there are $M$ RB in the whole system, and each RUE occupies one RB. F-APs and RRHs reuse the same RB set ${\rm{{\cal R}{\cal B}}} = \left\{ {1,2,...,M} \right\}$. Meanwhile, we assume that each F-AP occupies no more than one RB and serves two users simultaneously with NOMA technique. Furthermore, multiple F-APs can reuse the same RB to enhance the spectral efficiency. However, we need to set a threshold $z_{\max }$ to restrict the maximum number of F-APs occupying the same RB, which serves to efficiently restrain the co-channel interference.

Since different F-APs as well as F-AP and RRH may reuse the same RB, efficient RB allocation scheme is required to deal with the co-tier interference and the cross-tier interference. What's more, the two users served by one F-AP via NOMA protocol also demand an efficient power allocation scheme. For simplicity, the user association to F-APs and RRHs has been accomplished in advance. Therefore, in this paper, we focus on the resource allocation to maximize the sum rate of FUEs.

Provided that desired users have been scheduled, F-APs provide service for their users via NOMA. Therefore, the superposed signal at the $i$-th F-AP can be represented as:
\begin{eqnarray}
s_i  = \sqrt {\alpha _{i,1} P_i } s_{i,1}  + \sqrt {\alpha _{i,2} P_i } s_{i,2},
\end{eqnarray}
where $\alpha _{i,1}$ and $\alpha _{i,2}$ are power allocation coefficients, $\alpha _{i,1}  + \alpha _{i,2}  \le 1$ and $P_i$ is the transmit power of the $i$-th F-AP.

After the transmission, the received signal of the $n$-th $\left( {n = 1,2} \right)$ user served by $i$-th F-AP is:
\vspace{-1.5em}
\begin {spacing}{1.5}
\begin{eqnarray}
\begin{array}{l}
 y_{i,n}^m  = h_{i,n}^m \sqrt {\alpha _{i,1} P_i } s_{i,1}^m  + h_{i,n}^m \sqrt {\alpha _{i,2} P_i } s_{i,2}^m  + n_{i,n}^m  \\
 {\rm{    }}\ \ \ \ + \sum\nolimits_{\scriptstyle j \ne i \hfill \atop
  \scriptstyle j \in {\rm{{\cal F}{\cal A}{\cal P}}} \hfill} {\beta _{j,m} h_{j,i,n}^m \sqrt {P_j } s_j^m }  \\
  \quad\ \ + \sum\nolimits_{m \in {\rm{{\cal R}{\cal U}{\cal E}}}} {\beta _{i,m} l_{m,i,n} \sqrt {P_m } s_m }  \\
 \end{array}
\end{eqnarray}
\end {spacing}
\vspace{-0.5em}
where $h_{i,n}^m$, $h_{j,i,n}^m$ and $l_{m,i,n}$ represent the channel gain from $i$-th F-AP to its $n$-th user, from $j$-th F-AP to FUE $n$ served by $i$-th F-AP and from RRH $m$ to FUE $n$ served by $i$-th F-AP occupying RB $m$ respectively. $P_m$ is the transmit power of RRH $m$. $n_{i,n}^m$ is the additive white Gaussian noise (AWGN) with variance ${\sigma ^2 }$. $\beta _{i,m}$ is the RB allocation coefficient, with value equals 1 meaning that F-AP $i$ occupies RB $m$ and value equals 0 meaning that F-AP $i$ doesn't occupy RB $m$.

According to the NOMA principle and relative SIC mechanism to achieve resources multiplexing in power domain, assuming that user 1 served by F-AP $i$ on RB $m$ has to decode the signal for user 2 firstly and subtract it, then decode its own signal. While user 2 only needs to decode its own signal. The SINR for user 1 decoding its own signal is:
\begin{eqnarray}
SINR_{i,1}^m  = \frac{{\left| {h_{i,1}^m } \right|^2 \alpha _{i,1} P_i }}{{I_{co}^1  + I_{cr}^1  + \sigma ^2 }}
\end{eqnarray}

Similarly, the SINR for user 2 decoding its own signal is:
\begin{eqnarray}
SINR_{i,2}^m  = \frac{{\left| {h_{i,2}^m } \right|^2 \alpha _{i,2} P_i }}{{\left| {h_{i,2}^m } \right|^2 \alpha _{i,1} P_i  + I_{co}^2  + I_{cr}^2  + \sigma ^2 }}
\end{eqnarray}
where $I_{co}^1  = \sum\nolimits_{\scriptstyle j \ne i \hfill \atop
  \scriptstyle j \in {\rm{{\cal F}{\cal A}{\cal P}}} \hfill} {\beta _{j,m} \left| {h_{j,i,1}^m } \right|^2 P_j }$ represents the co-tier interference from other F-APs occupying the same RB $m$ to user 1 served by F-AP $i$ and $I_{co}^2  = \sum\nolimits_{\scriptstyle j \ne i \hfill \atop
  \scriptstyle j \in {\rm{{\cal F}{\cal A}{\cal P}}} \hfill} {\beta _{j,m} \left| {h_{j,i,2}^m } \right|^2 P_j }$ represents the co-tier interference from other F-APs occupying the same RB $m$ to user 2 served by F-AP $i$. $I_{cr}^1  = \sum\nolimits_{m \in {\rm{{\cal R}{\cal U}{\cal E}}}} {\beta _{i,m} \left| {l_{m,i,1} } \right|^2 P_m }$ and $I_{cr}^2  = \sum\nolimits_{m \in {\rm{{\cal R}{\cal U}{\cal E}}}} {\beta _{i,m} \left| {l_{m,i,2} } \right|^2 P_m }$ represents the cross-tier interference from RRH reusing the same RB $m$.

Based on the discussion above, data rates of two users served by F-AP $i$ on RB $m$ are given as follows:
\begin{eqnarray}
R_{i,1}^m  = \beta _{i,m} B\log _2 \left( {1 + SINR_{i,1}^m } \right),
\end{eqnarray}
\vspace{-1.5em}
\begin{eqnarray}
R_{i,2}^m  = \beta _{i,m} B\log _2 \left( {1 + SINR_{i,2}^m } \right),
\end{eqnarray}

\vspace{0.1em}
\section{Problem Formulation}
\vspace{0.3em}
Given that the user association to F-APs and RRHs has been accomplished before, in this section, we only focus on the users under the service of F-APs and optimize the resource allocation scheme, including RB allocation and power allocation to maximize the sum rate of FUEs.

In order to avoid the situation that one user under F-AP $i$ occupies all resources of RB $m$, we introduce weights $w_{i,n}$ [7] and formulate the weighted sum rate maximization problem as follows:
\begin{eqnarray}
\begin{array}{l}
 {}_{{\bm{\beta }},{\bm{\alpha }}}^{\max } {\rm{   }}\qquad R_{sum}  = \sum\limits_{i = 1}^N {\sum\limits_{m = 1}^M {w_{i,1}R_{i,1}^m  + w_{i,2}R_{i,2}^m } }  \\
 s.t.{\rm{   }}\qquad\beta _{i,m}  \in \left\{ {0,1} \right\},\forall i \in {\rm{{\cal F}{\cal A}{\cal P}}},\forall m \in {\rm{{\cal R}{\cal B}}},(a) \\
 {\rm{      }}\qquad\quad\sum\limits_{m = 1}^M {\beta _{i,m}  \le 1,} \forall i \in {\rm{{\cal F}{\cal A}{\cal P}}},\qquad\ \ \ \ \ \ \ \ \ (b) \\
 {\rm{      }}\quad\ \ \ \ \ \ \sum\limits_{i = 1}^N {\beta _{i,m}  \le z_{\max } ,} \forall m \in {\rm{{\cal R}{\cal B}}},\qquad\qquad\ (c) \\
\quad \quad\quad\ \sum\limits_{i = 1}^N {\sum\limits_{m = 1}^M {\beta _{i,m} P_i \left| {g_{i,m} } \right|^2 } }  \le I_{\max }^m ,\quad\ \  \ \ \ \ (d)\\
 {\rm{      }}\qquad\quad\ \alpha _{i,1}  \ge 0,{\rm{ }}\alpha _{i,2}  \ge 0,{\rm{ }}\forall i \in {\cal N},{\rm{       }}\qquad\qquad\ \ (e)\ \  \\
 {\rm{      }}\qquad\quad\ \alpha _{i,1}  + \alpha _{i,2}  \le 1,\forall i \in {\cal N},{\rm{             }}\qquad\qquad\ \ \ \ \ (f) \\
 \end{array}
\end{eqnarray}
where ${\bm{\beta }} = \left\{ {\beta _{i,m} |\forall i \in {\rm{{\cal F}{\cal A}{\cal P}}},\forall m \in {\rm{{\cal R}{\cal B}}}} \right\}$ and ${\bm{\alpha }} = \left\{ {\alpha _{1i} ,\alpha _{2i} |i \in {\rm{{\cal F}{\cal A}{\cal P}}}} \right\}$ are RB allocation matrix and power allocation vector respectively. $I_m=\sum\limits_{i = 1}^N {\sum\limits_{m = 1}^M {\beta _{i,m} P_i \left| {g_{i,m} } \right|^2 } }$ is the interference of RUE $m$ from FUEs which reuse the same RB $m$ and $g_{i,m}$ is the channel coefficient from F-AP $i$ to RUE $m$ on RB $m$. Constraint (b) represents that each F-AP can only occupy no more than one RB. Constraint (c) is used to restrain the maximum number of F-APs reusing the same RB $m$ to relieve the co-channel interference. Constraint (d) guarantees that the whole interference from FUEs on same RB $m$ won't exceed $I_{\max }^m$, which is the maximum interference threshold of RUE $m$. Constraint (e) means that the transmit power of users in F-AP $i$ is required to be non-negative and constraint (f) gives the upper bound of the transmit power of each NOMA-pair users.


\section{Proposed Efficient Solution Methodology}
\vspace{0.2em}
Due to the discrete binary variables of RB allocation scheme and the non-convexity of the objective function, the optimization problem in (7) is a mixed non-convex problem. To solve this problem, we need to firstly decouple the original problem into a RB allocation problem and a power allocation problem. Given that power allocation scheme has been confirmed, the RB allocation for F-APs is formulated as a many-to-one matching game [8], in which multiple F-APs interact with one RB to find the optimal matching. Similarly, given that RB allocation scheme has been determined, we can rewrite the power allocation problem as a difference of convex function. Some approximations have been used to convert it into a convex one and we get the closed-form expression of the optimal power allocation coefficients. Then successive convex programming is utilized to tighten the upper bound until convergence. Finally, we combine the two above to propose an iterative algorithm to get the optimal joint resource allocation scheme.

\subsection{RB Allocation Scheme}
Under the circumstance that power allocation scheme has been determined, the original problem can be reformulated as follows:
\begin{eqnarray}
\begin{array}{l}
 {}_{\bm{\beta }}^{\max } {\rm{   }}\qquad R_{sum}  \\
 s.t.{\rm{   }}\qquad \beta _{i,m}  \in \left\{ {0,1} \right\},\forall i \in {\rm{{\cal F}{\cal A}{\cal P}}},\forall m \in {\rm{{\cal R}{\cal B}}},(a) \\
 {\rm{      }}\qquad\ \ \ \sum\limits_{m = 1}^M {\beta _{i,m}  \le 1,} \forall i \in {\rm{{\cal F}{\cal A}{\cal P}}},\qquad\quad  \ \ \ \ \ \ (b) \\
 {\rm{      }}\qquad\ \ \ \sum\limits_{i = 1}^N {\beta _{i,m}  \le z_{\max } ,} \forall m \in {\rm{{\cal R}{\cal B}}},\qquad \quad\ \ \ \ (c) \\
\qquad \ \ \ \sum\limits_{i = 1}^N {\sum\limits_{m = 1}^M {\beta _{i,m} P_i \left| {g_{i,m} } \right|^2 } }  \le I_{\max }^m ,\qquad \ \ \ (d)
 \end{array}
\end{eqnarray}

To solve the problem above, we take advantage of matching method and formulate it as a many-to-one matching problem. We define $\Phi$ as a matching function with $\left| {\Phi \left( i \right)} \right| = 1,\forall i \in {\rm{{\cal F}{\cal A}{\cal P}}}$ means that each F-AP can only be matched with one RB, and $\left| {\Phi \left( m \right)} \right| \le z_{\max } ,\forall m \in {\rm{{\cal R}{\cal B}}}$ means that the maximum number of F-APs matching with same RB $m$ cannot exceeds the threshold $z_{\max }$.

The utility function of F-AP $i$ and RB $m$ can be expressed as follows respectively:
\begin{eqnarray}
U_i  = \left( {w_{i,1}R_{i,1}^m  + w_{i,2}R_{i,2}^m } \right)
\end{eqnarray}
\vspace{-1.8em}
\begin{eqnarray}
U_m  = \sum\limits_{i = 1}^N {\beta _{i,m} \left( {w_{i,1}R_{i,1}^m  + w_{i,2}R_{i,2}^m } \right)}
\end{eqnarray}
where $U_i$ represents the weighted sum rate of users served by F-AP $i$ on RB m, and $U_m$ represents the weighted sum rate of all the serving users occupying RB m. Both of them are related to the objective function.

\begin{algorithm}[!htbp]

\caption{Modified Swap-enabled Matching Algorithm (MSEMA)}
\label{alg:Framwork}
\begin{algorithmic}[1]
\STATE Form ${\rm{{\cal F}}}_i$ and ${\rm{{\cal R}}}_m$ of all the F-APs and RBs respectively;\\
\STATE Form the set ${\rm{{\cal U}}}$ including all the F-APs that are not matched;\\
\STATE \textbf{while} ${\rm{{\cal U}}} \ne \emptyset$ \textbf{do}\\
\STATE ~ \textbf{for} $\forall i \in {\rm{{\cal F}{\cal A}{\cal P}}}$ \textbf{do}\\
\STATE ~ ~ F-AP $i$ matches to its most preferred RB in ${\rm{{\cal F}}}_i$ that hasn't reject it before;\\
\STATE ~ \textbf{end for}
\STATE ~ \textbf{for} $\forall m \in {\rm{{\cal R}{\cal B}}}$ \textbf{do}\\
\STATE ~ ~ \textbf{if} $\sum\limits_{i = 1}^N {\beta _{i,m} }  \le z_{\max }$ \textbf{then}\\
\STATE ~ ~ ~ RB $m$ holds all the matched F-APs and remove them from ${\rm{{\cal U}}}$;\\
\STATE ~ ~ \textbf{else}
\STATE ~ ~ ~ RB $m$ holds the $z_{\max }$ F-APs in ${\rm{{\cal R}}}_m$ and remove them from ${\rm{{\cal U}}}$. Then reject the others;\\
\STATE ~ ~ \textbf{end if}
\STATE ~ ~ Delete $m$ from ${\rm{{\cal F}}}_i$ that has sent proposals;
\STATE ~ \textbf{end for}
\STATE \textbf{end while}
\STATE Set ${\rm{{\cal S}}}_{i,j}  = 0$;
\STATE \textbf{while} there exists swap-pair \textbf{do}\\
\STATE ~ \textbf{if} F-AP $i$ and $j$ construct a swap-pair and ${\rm{{\cal S}}}_{i,j}  + {\rm{{\cal S}}}_{j,i}  < 2$ \textbf{then}\\
\STATE ~ ~ Update $\Phi _i^j$ and ${\rm{{\cal S}}}_{i,j}  = {\rm{{\cal S}}}_{i,j}  + 1$;
\STATE ~ \textbf{else}
\STATE ~ ~ $\Phi _i^j$ remains the same;
\STATE ~ \textbf{end if}
\STATE \textbf{end while}
\end{algorithmic}
\end{algorithm}

In the matching process, both F-APs and RBs are firstly required to determine their preference list according to their own interests. For simplicity, we assume that the preference list is set in a descending order. To be more specific, with the utility function on all the RBs, F-AP $i$ is able to determine its own preference list ${\rm{{\cal F}}}_i$. Similarly, with the utility function over all possible sets of F-APs, RB $m$ can also set its own preference list ${\rm{{\cal R}}}_m$. It's noteworthy that multiple F-APs may occupy the same RB and bring about peer effects in the matching process, which results in variations of preference lists with matching goes on. Hence, we propose a modified swap-enabled matching algorithm to solve all these problems above.

$\Phi _i^j  = \left\{ {\left( {i,\Phi \left( j \right)} \right),\left( {j,\Phi \left( i \right)} \right)} \right\} \cup \left\{ {\Phi \backslash \left\{ {\left( {i,\Phi \left( i \right)} \right),\left( {j,\Phi \left( j \right)} \right)} \right\}} \right\}$ is defined as the swap matching function, in which only F-AP $i$ and $j$ switch. Compared with traditional swap operations in [9], which performs the swap if and only if the condition $U_s \left( {\Phi _i^j } \right) \ge U_s \left( \Phi  \right),\forall s \in \left\{ {i,j,\Phi _i ,\Phi _j } \right\}$ is satisfied, we propose a modified swap-enabled matching scheme. Since that M RUEs have been allocated to M RBs ahead of time, the interference to each RUE needs attention. We take interference to RUE $m$ caused by other FUEs occupying the same RB $m$ into consideration and define the modified swap-pair as follows:

\textit{Definition 1: F-AP i and j is a swap-pair if and only if}
\vspace{-1em}
\begin{spacing}{1.2}
\begin{eqnarray}
\begin{array}{l}
 a.\ I_{tol}^{\Phi \left( i \right)}  + I_i^{\Phi \left( i \right)}  - I_j^{\Phi \left( i \right)}  \ge 0, \\
 b.\ I_{tol}^{\Phi \left( j \right)}  + I_j^{\Phi \left( j \right)}  - I_i^{\Phi \left( j \right)}  \ge 0, \\
 c.\ U_s \left( {\Phi _i^j } \right) \ge U_s \left( \Phi  \right),\forall s \in \left\{ {i,j,\Phi _i ,\Phi _j } \right\},
 \end{array}
\end{eqnarray}
\end{spacing}
\vspace{-0.5em}
\textit{where $I_{tol}^m  = I_{\max }^m  - I_m$ is the residual interference that can be tolerated by subchannel $m$.}

The requirements $a$ and $b$ correspond to the constraint (7d), which mean that if the changed interference caused by swap operation is larger than the residual interference that can be tolerated, the swap won't happen. The requirement $c$ means that the swap operation should be beneficial to the utility function of both F-AP and RB.

In this algorithm, to prevent meaningless swap operations between two F-APs $i$ and $j$, we define a variable ${\rm{{\cal S}}}_{i,j}$ to count the times of F-AP $i$ and $j$ swap their former matched RBs.

\subsection{Power Allocation Scheme}
With given RB allocation scheme, the optimization problem in (7) is reduced to:
\begin{eqnarray}
\begin{array}{l}
 {}_{\bm{\alpha }}^{\max } {\rm{   }}\qquad R_{sum}  = \sum\limits_{i = 1}^N {\sum\limits_{m = 1}^M {w_{i,1}R_{i,1}^m  + w_{i,2}R_{i,2}^m } }  \\
 s.t.\qquad {\rm{   }}(7e),(7f) \\
 \end{array}
\end{eqnarray}

Since the power allocation of each F-AP is independent, the problem of maximizing the weighted sum rate of all FUEs is equal to the problem of maximizing the weighted sum rate of each F-AP. Therefore, in this section, the original optimization problem can be simplified to the power allocation problem for the $i$-th F-AP as follows:
\begin{eqnarray}
\begin{array}{l}
 {}_{\left\{ {\alpha _{i,1} ,\alpha _{i,2} |i \in {\rm{{\cal N}}}} \right\}}^{\qquad \max } {\rm{        }}\quad w_{i,1}R_{i,1}^m  + w_{i,2}R_{i,2}^m   \\
 \qquad s.t.\qquad\quad (7e)(7f)
 \end{array}
\end{eqnarray}

The optimization problem in (13) is a non-convex problem as a result of the co-channel interference. Therefore, in this section, successive convex programming [10] is exploited to deal with the power allocation problem of each NOMA-pair and obtain a locally optimal solution of (13).

We can see that the objection function in (13) is non-convex, which can be rewritten as:
\begin{eqnarray}
\begin{array}{l}
{_{\left\{ {\alpha _{i,1} ,\alpha _{i,2} |i \in N} \right\}}^{\qquad \min } \quad  - \left( {F\left( {{\bm{\alpha }}_i } \right) - G\left( {{\bm{\alpha }}_i } \right)} \right)}  \\
 \qquad s.t.\qquad\qquad (7e)(7f)
 \end{array}
\end{eqnarray}
$F\left( {{\bm{\alpha }}_i } \right)$ and $G\left( {{\bm{\alpha }}_i } \right)$ are defined in (15) and (16) concretely.
\begin{eqnarray}
\begin{array}{l}
 F\left( {{\bm{\alpha }}_i } \right) = w_{i,1} B\log _2 \left( {c_1  + \left| {h_{i,1} } \right|^2 \alpha _{i,1} P_i } \right) \\
  \qquad\quad+ w_{i,2} B\log _2 \left( {\left| {h_{i,2} } \right|^2 \left( {\alpha _{i,1}  + \alpha _{i,2} } \right)P_i  + c_2 } \right), \\
 \end{array}
\end{eqnarray}
\vspace{-0.5em}
\begin{eqnarray}
\begin{array}{l}
 G\left( {{\bm{\alpha }}_i } \right) = w_{i,1} B\log _2 \left( {c_1 } \right) \\
  \qquad\quad+ w_{i,2} B\log _2 \left( {\left| {h_{i,2} } \right|^2 \alpha _{i,1} P_i  + c_2 } \right), \\
 \end{array}
\end{eqnarray}
where $c_1  = I_{co}^1  + I_{cr}^1  + \sigma ^2$ and $c_2  = I_{co}^2  + I_{cr}^2  + \sigma ^2$.
\begin{spacing}{1}
It's easy to find that $F\left( {{\bm{\alpha }}_i } \right)$ and $G\left( {{\bm{\alpha }}_i } \right)$ are both convex functions. Therefore, the problem in (14) is a canonical form of difference of convex (D.C.) function programming [11]. Since $G\left( {{\bm{\alpha }}_i } \right)$ is a differentiable function, for any feasible solution ${\bm{\tilde \alpha }}_i$, we can get the following inequality by use of first-order Taylor expansion:
\end{spacing}
\vspace{-0.5em}
\begin{small}
\begin{eqnarray}
G\left( {{\bm{\alpha }}_i } \right) \ge G\left( {{\bm{\tilde \alpha }}_i } \right) + \nabla _{{\bm{\alpha }}_i } G\left( {{\bf{\tilde \alpha }}_i } \right)\left( {{\bm{\alpha }}_i  - {\bm{\tilde \alpha }}_i } \right) \buildrel \Delta \over = \bar G\left( {{\bm{\alpha }}_i ,{\bm{\tilde \alpha }}_i } \right),
\end{eqnarray}
\end{small}
\noindent where $\nabla _{{\bm{\alpha }}_i } G\left( {{\bm{\tilde \alpha }}_i } \right)$ is the gradient of $G\left( {{\bm{\alpha }}_i } \right)$ at a given point ${{\bm{\tilde \alpha }}_i }$, $\bar G\left( {{\bm{\alpha }}_i ,{\bm{\tilde \alpha }}_i } \right)$ is the affine function and represents the global estimation of $G\left( {{\bm{\alpha }}_i } \right)$.
\vspace{-0.5em}
\begin{algorithm}[!htbp]
\renewcommand{\algorithmicrequire}{\textbf{Input:}}
\renewcommand\algorithmicensure {\textbf{Output:}}
\caption{Iterative Algorithm for Power Allocation}
\label{alg:Framwork}
\begin{algorithmic}[1]
\REQUIRE ~\\
${\bm{\alpha }}_i$: power allocation vector;\\
$\varepsilon _{thr}$: a threshold\\
$J$: maximum iteration number
\ENSURE ~\\
${\bm{\alpha }}_i$: power allocation vector;\\

\STATE \textbf{Initialization}:\\
 $j = 0,{\bm{\alpha }}_i  = {\bm{\alpha }}_i^{(0)};$\\
\STATE \textbf{while} $\left| {H({\bm{\alpha }}_i^{(j)} ) - H({\bm{\alpha }}_i^{(j - 1)} )} \right| \ge \varepsilon _{thr}$ and $j<J$ \textbf{do}\\
\STATE ~ $j=j+1$;
\STATE ~ ${\bm{\tilde \alpha }}_i^{(j)}  = {\bm{\alpha }}_i^{(j - 1)}$;\\
\STATE ~ solve the convex optimization problem in (18) to get the \\~ power allocation vector ${\bm{\alpha }}_i^{(j)}$;\\
\STATE ~ compute $H({\bm{\alpha }}_i^{(j)} )$ according to the solution ${\bm{\alpha }}_i^{(j)}$ obtained\\ ~ in the previous step;\\
\STATE \textbf{end} \textbf{while}\\
\STATE output the power allocation vector: ${\bm{\alpha }}_i^*  = {\bm{\alpha }}_i^{(j)}$;\\
\end{algorithmic}
\end{algorithm}
\vspace{-1em}

Therefore, we get the following optimization problem, which is a convex one and provides an upper bound for the problem (13) according to (16).
\begin{eqnarray}
\begin{array}{l}
   {_{\left\{ {\alpha _{i,1} ,\alpha _{i,2} |i \in N} \right\}}^{\qquad \min } \quad H\left( {{\bm{\alpha }}_i } \right)= - F\left( {{\bm{\alpha }}_i } \right) + \bar G\left( {{\bm{\alpha }}_i ,{\bm{\tilde \alpha }}_i } \right)}  \\
   {\qquad s.t.\qquad \qquad \alpha _{i,1}  + \alpha _{i,2}  \le 1}  \\
\end{array}
\end{eqnarray}
where we have
\vspace{-0.5em}
\begin{eqnarray}
\bar G\left( {{\bm{\alpha }}_i ,{\bm{\tilde \alpha }}_i } \right) = G\left( {{\bm{\tilde \alpha }}_i } \right) + \left( {{\bf{\alpha }}_{i,1}  - \tilde \alpha _{i,1} } \right)\frac{{w_{i,2} B\left| {h_{i,2} } \right|^2 P_i }}{{\left| {h_{i,2} } \right|^2 \tilde \alpha _{i,1} P_i  + c_2 }}.
\end{eqnarray}
\begin{eqnarray}
\begin{array}{l}
 G\left( {{\bm{\tilde \alpha }}_i } \right) = w_{i,1} B\log _2 \left( {c_1 } \right) \\
  \qquad\quad+ w_{i,2} B\log _2 \left( {\left| {h_{i,2} } \right|^2 \tilde \alpha _{i,1} P_i  + c_2 } \right) \\
 \end{array}
\end{eqnarray}

Then we are able to iteratively update the power allocation vector ${\bm{\alpha }}_i  = \left[ {\alpha _{i,1} ,\alpha _{i,2} } \right]$ by solving the optimization problem in (18) to tighten the upper bound until convergence. The power allocation algorithm using successive convex programming is shown in Algorithm 2. Firstly, we initialize the power allocation vector randomly and then perform the iteration to update it. In the $j$-th iteration, we set ${\bm{\tilde \alpha }}_i^{(j)}  = {\bm{\alpha }}_i^{(j - 1)}$ and get ${\bm{\alpha }}_i $ by solving the convex optimization problem in (18). The circulation continues until convergence or the number of iterations exceeds the max number of iterations $J$.

The convex optimization problem in (18) is solved by utilizing the Lagrange multiplier method and Karush-Kuhn-Tucher (KKT) conditions. We set up the Lagrange function of problem (18) as:
\begin{eqnarray}
L\left( {{\bm{\alpha }}_i ,\lambda } \right) =  - F\left( {{\bm{\alpha }}_i } \right) + \bar G\left( {{\bm{\alpha }}_i ,{\bm{\tilde \alpha }}_i } \right) + \lambda \left( {\alpha _{i,1}  + \alpha _{i,2}  - 1} \right)
\end{eqnarray}
where $\lambda$ is the Lagrange multiplier.

We put (15)(19) into $L\left( {{\bm{\alpha }}_i ,\lambda } \right)$ and apply KKT conditions to find the optimal solution as follows:
\vspace{-1.5em}
\begin{spacing}{1.5}
\begin{eqnarray}
\left\{ \begin{array}{l}
  - \frac{{w_{i,1} B\left| {h_{i,1} } \right|^2 P_i }}{{c_1  + \left| {h_{i,1} } \right|^2 \alpha _{i,1} P_i }} - \frac{{w_{i,2} B\left| {h_{i,2} } \right|^2 P_i }}{{\left| {h_{i,2} } \right|^2 \alpha _{i,1} P_i  + c_2  + \left| {h_{i,2} } \right|^2 \alpha _{i,2} P_i }} \\ \qquad\quad\qquad\qquad+ \frac{{w_{i,2} B\left| {h_{i,2} } \right|^2 P_i }}{{\left| {h_{i,2} } \right|^2 \tilde \alpha _{i,1} P_i  + c_2 }} + \lambda \ln 2 = 0 \\
  - \frac{{w_{i,2} B\left| {h_{i,2} } \right|^2 P_i }}{{\left| {h_{i,2} } \right|^2 \alpha _{i,1} P_i  + c_2  + \left| {h_{i,2} } \right|^2 \alpha _{i,2} P_i }} + \lambda \ln 2 = 0 \\
 \lambda \left( {\alpha _{i,1}  + \alpha _{i,2}  - 1} \right) = 0 \\
 \alpha _{i,1}  + \alpha _{i,2}  \le 1 \\
 \end{array} \right.
\end{eqnarray}
\end{spacing}
\vspace{-0.8em}
We can show that $\lambda  \ne 0$, because if $\lambda = 0$, the KKT conditions lead to the following two equations:
\vspace{-1.5em}
\begin{spacing}{1.5}
\begin{eqnarray}
\left\{ \begin{array}{l}
  - \frac{{w_{i,1} B\left| {h_{i,1} } \right|^2 P_i }}{{c_1  + \left| {h_{i,1} } \right|^2 \alpha _{i,1} P_i }} - \frac{{w_{i,2} B\left| {h_{i,2} } \right|^2 P_i }}{{\left| {h_{i,2} } \right|^2 \alpha _{i,1} P_i  + c_2  + \left| {h_{i,2} } \right|^2 \alpha _{i,2} P_i }} \\ \qquad\qquad\qquad\quad+ \frac{{w_{i,2} B\left| {h_{i,2} } \right|^2 P_i }}{{\left| {h_{i,2} } \right|^2 \tilde \alpha _{i,1} P_i  + c_2 }} = 0 \\
  - \frac{{w_{i,2} B\left| {h_{i,2} } \right|^2 P_i }}{{\left| {h_{i,2} } \right|^2 \alpha _{i,1} P_i  + c_2  + \left| {h_{i,2} } \right|^2 \alpha _{i,2} P_i }} = 0 \\
 \end{array} \right.
\end{eqnarray}
\end{spacing}
\vspace{-0.5em}
\noindent which cannot be true. Therefore, we have $\alpha _{i,1}  + \alpha _{i,2}  - 1 = 0$.

With some algebraic manipulations, the optimal solutions of closed form can be obtained as follows:
\begin{eqnarray}
\alpha _{i,1}  = \frac{{w_{i,1} \left( {\left| {h_{i,2} } \right|^2 \tilde \alpha _{i,1} P_i  + c_2 } \right)}}{{w_{i,2} \left| {h_{i,2} } \right|^2 P_i }} - \frac{{c_1 }}{{\left| {h_{i,1} } \right|^2 P_i }}
\end{eqnarray}
\vspace{-2em}
\begin{spacing}{1.3}
\begin{eqnarray}
\begin{array}{l}
 \alpha _{i,2}  = \frac{{w_{i,2}B }}{{\lambda \ln 2}} - \frac{{w_{i,1} \left( {\left| {h_{i,2} } \right|^2 \tilde \alpha _{i,1} P_i  + c_2 } \right)}}{{w_{i,2} \left| {h_{i,2} } \right|^2 P}} \\
  \quad\ \ + \frac{{c_1 }}{{\left| {h_{i,1} } \right|^2 P_i }} - \frac{{c_2 }}{{\left| {h_{i,2} } \right|^2 P_i }} \\
 \end{array}
\end{eqnarray}
\end{spacing}
\noindent where $\lambda$ can be obtained by putting (24) and (25) into $\alpha _{i,1}  + \alpha _{i,2}  - 1 = 0$.
\begin{eqnarray}
\lambda  = \frac{{w_{i,2} B\left| {h_{i,2} } \right|^2 P_i }}{{\ln 2\left( {c_2  + \left| {h_{i,2} } \right|^2 P_i } \right)}}
\end{eqnarray}

\subsection{Joint RB and Power Allocation}
Based on RB allocation scheme and power allocation scheme discussed above, we propose a joint resource allocation algorithm.

In Algorithm 3, we randomly initialize the power allocation of all NOMA users served by F-APs. Then, for the given power allocation scheme, it's easy to exploit Algorithm 1 to update RB allocation scheme. Similarly, for the given RB allocation scheme obtained in the previous step, we are able to update power allocation result according to Algorithm 2. Such operations are performed iteratively until converges to a stationary point.
\begin{algorithm}[!htbp]
\renewcommand{\algorithmicrequire}{\textbf{Input:}}
\renewcommand\algorithmicensure {\textbf{Output:}}
\caption{Joint Resource Allocation Algorithm}
\label{alg:Framwork}
\begin{algorithmic}[1]
\REQUIRE ~\\
${\bm{\alpha }}$: power allocation coefficients vector;\\
$x_{\max }$: max number of iterations;\\
\ENSURE ~\\
${\bm{\alpha }}$: power allocation coefficients vector;\\
$\bm{\beta}$: RB allocation matrix.

\STATE \textbf{Initialization}:\\
$x = 0,\bm{\alpha}  = \bm{\alpha ^{(0)}};$\\
\STATE \textbf{while} $x<x_{\max}$ \textbf{do}\\
\STATE ~ Update RB allocation $\bm{\beta}$ according to Algorithm 1;
\STATE ~ Update power allocation $\bm{\alpha}$ according to Algorithm 2;\\
\STATE ~ $x=x+1$;\\
\STATE ~ until convergence;\\
\STATE \textbf{end} \textbf{while}\\
\end{algorithmic}
\end{algorithm}

\vspace{-0.1em}

\section{Simulation Results And Analysis}
In this section, we investigate the performance of the proposed algorithms. We consider a NOMA-enabled downlink F-RAN with $N=6$ F-APs, $M=3$ RRHs and propose resource allocation methods to maximize the weighted sum rate of FUEs with fixed $w_{i,1}=0.9,w_{i,2}=1.1,\forall i \in {\rm{{\cal F}{\cal A}{\cal P}}}$. The transmit power is 23 dBm for each F-AP and 13 dBm for each RRH. The noise power spectral density is -174 dBm/Hz and the path-loss exponent is set as 4.

\begin{figure}[htbp]
\centering
\includegraphics[width=0.37\textwidth]{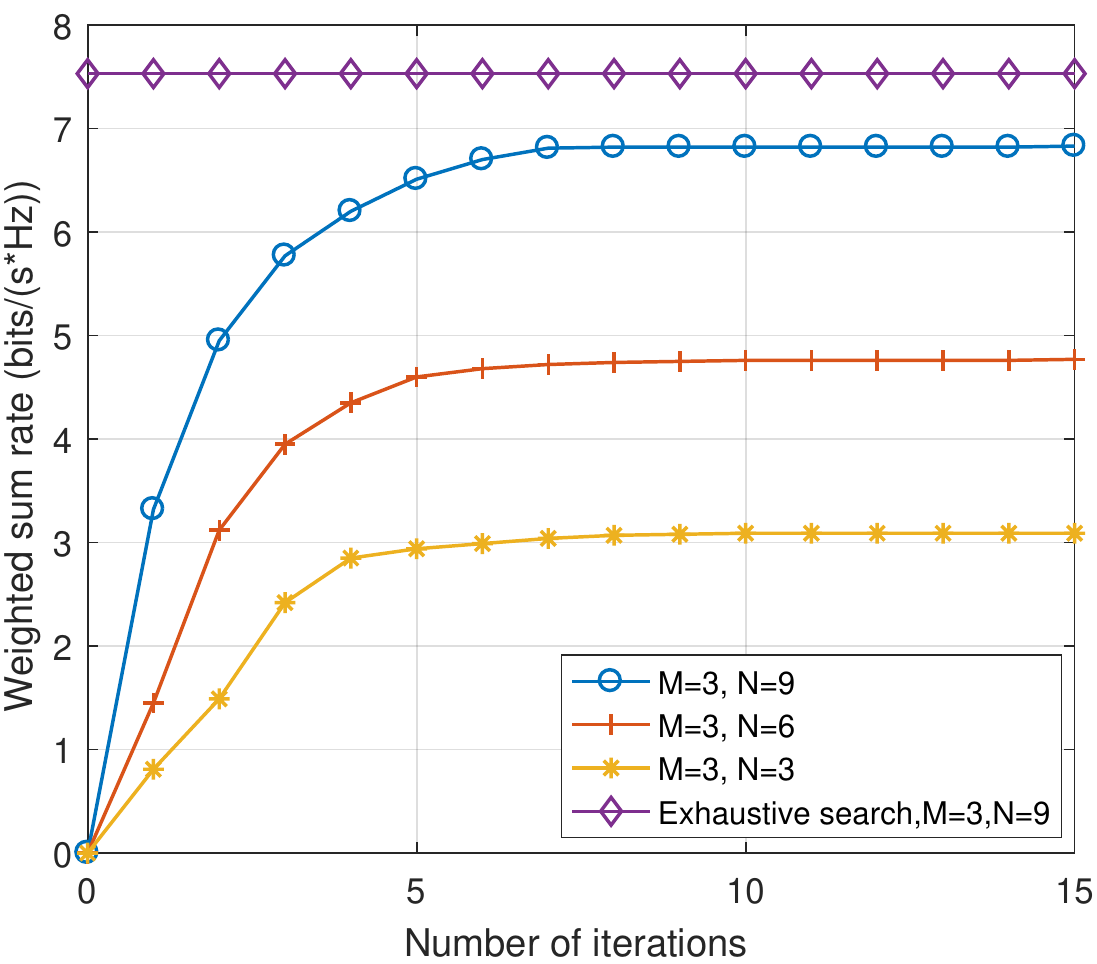}
\caption{Convergence behavior under different number of F-APs, $M=3$.}
\end{figure}
The convergence of algorithm under different number of F-APs with 3 RBs is shown in fig.2. It's easy to see that Algorithm 3 converges to a stationary point in about 5-10 iterations and the number of iterations to converge increases as the number of F-APs grows. The reason is that with more F-APs, the matching choice between F-APs and certain RB also increases. Besides, we can also see that performance of the proposed algorithm is very close to exhaustive searching algorithm. What's more, by comparing these three lines of different $N$, we can see that the growth of $N$ contributes to the higher sum rate.

\begin{figure}[htbp]
\centering
\includegraphics[width=0.37\textwidth]{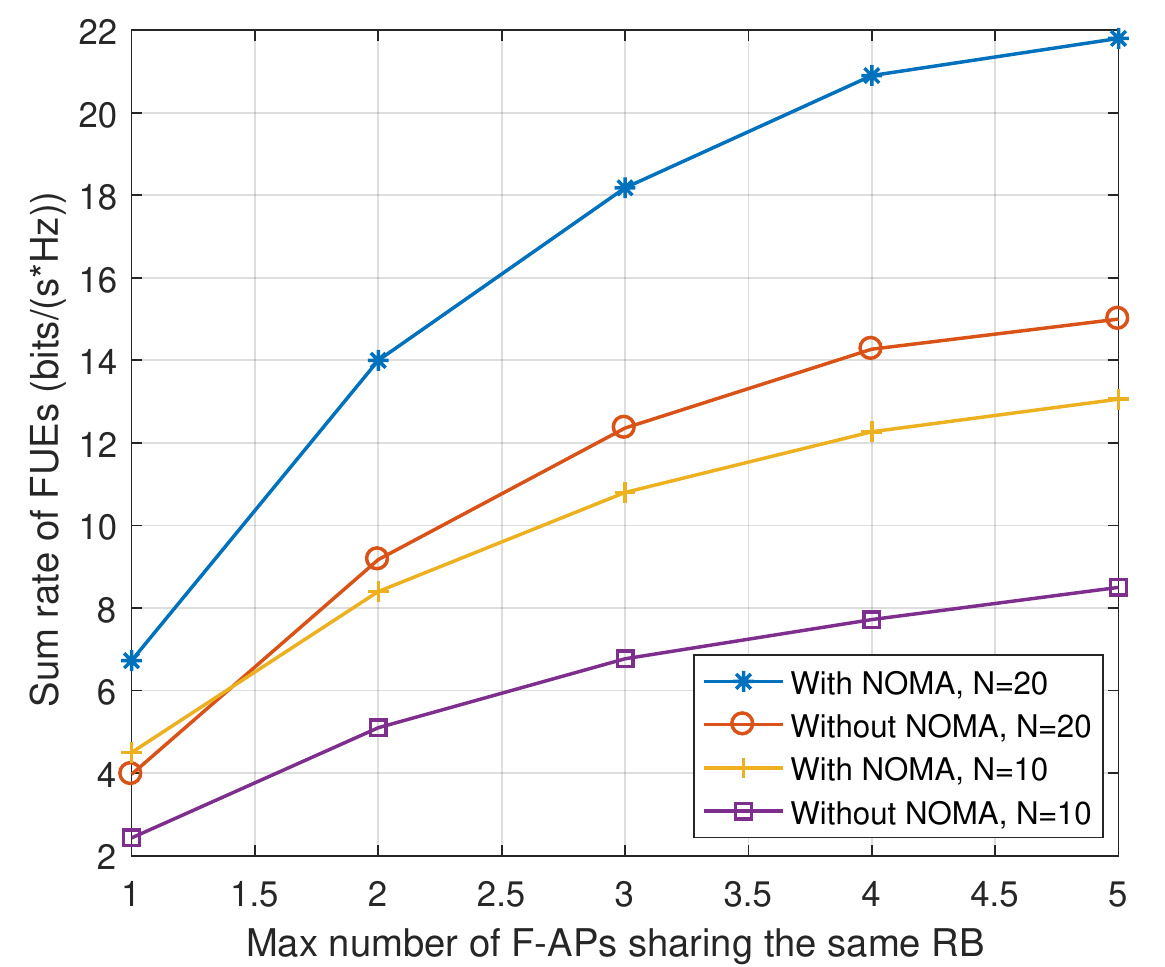}
\caption{Sum rate of NOMA-pairs with various maximum number of F-APs occupying the same RB, $M=3$.}
\end{figure}

For Fig.3, the relationship between the maximum number of F-APs occupying the same RB and sum rate of FUEs is provided. It's observed that the sum rate increases dramatically when $z_{max}$ is relatively small and increases slower and slower. Furthermore, we compare the sum rate of system using NOMA technology with the sum rate of system not using NOMA. It's easy to find that the use of NOMA technology increases the sum rate overwhelmingly. Meanwhile, we can also see that sum rate with NOMA is smaller than twice of sum rate without NOMA, which owing to the co-tier interference from other F-APs and the cross-tier interference from RRH occupying the same RB.
\begin{figure}[htbp]
\centering
\includegraphics[width=0.37\textwidth]{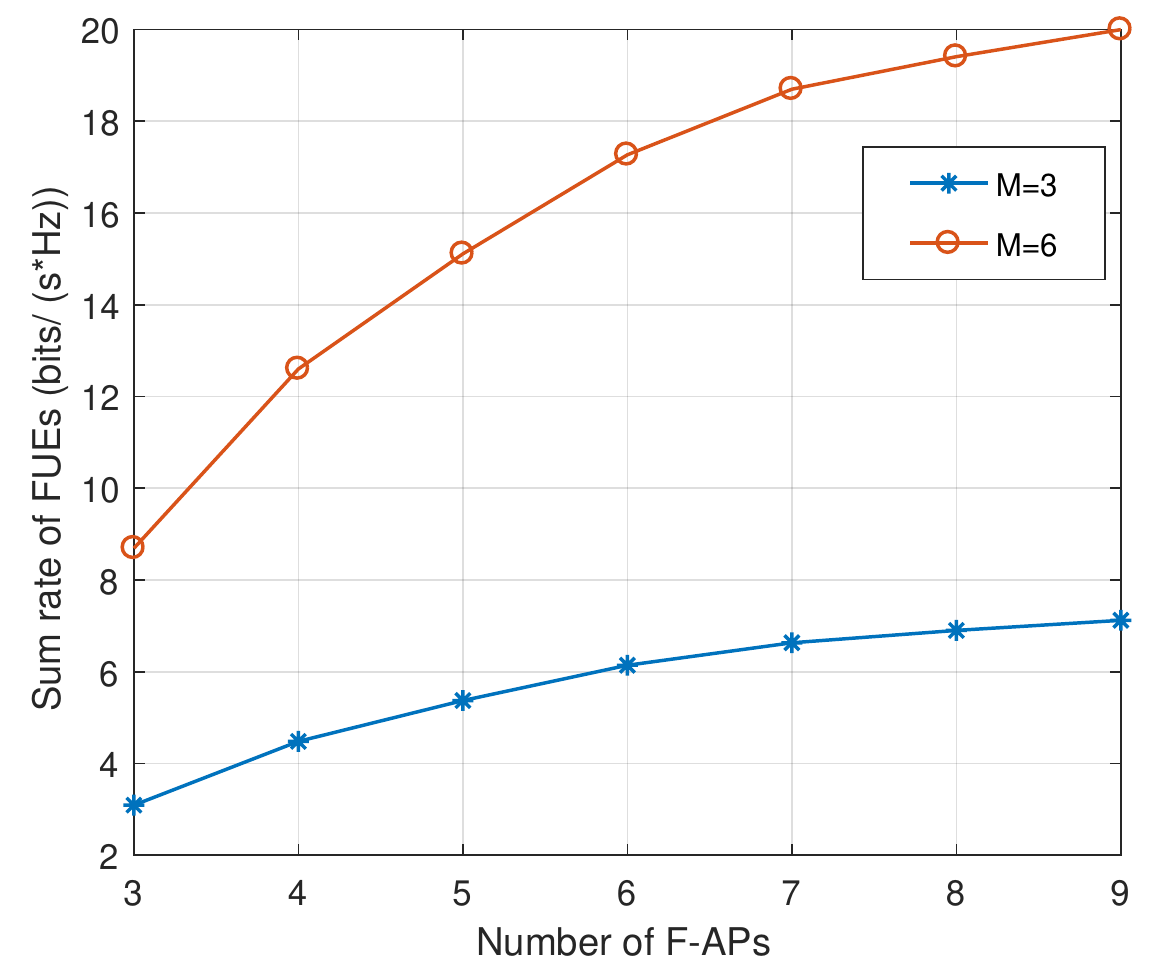}
\caption{Sum rate of NOMA-pairs versus number of F-APs.}
\end{figure}

Fig.4 depicts the impact of F-APs number on sum rate of FUEs, under the case of $M=3$ and $M=6$. we can obviously find that as the number of F-APs grows, the sum rate of F-UEs increases monotonically. It's also worth noting that the increase of RBs number serves to enhance the sum rate enormously. The reason is that with more RBs, more F-APs can be matched to them through the modified swap-enabled matching algorithm, which contributes to the sum rate.
\vspace{0.5em}
\section{Conclusion}
In this paper, we investigate the resource allocation problem of a NOMA-enabled downlink F-RAN, in which NOMA technique is utilized to enhance the spectral efficiency of the system. With the aim to maximize the sum rate of FUEs, we formulate an optimization problem about resource allocation and decouple it into a RB allocation problem and a power allocation problem. Being modeled as a many-to-one matching problem, the former question is solved by a modified swap-enabled matching algorithm. With a given RB allocation scheme, we exploit some approximations to convert the power allocation problem, which is non-convex, into a tractable one. Performance of the proposed algorithm is illustrated via numerical results and it can be verified that the use of NOMA technique outperforms the conventional OMA scenarios.

\section{ACKNOWLEDGEMENT}
This work was supported in part by the National Natural Science Foundation of China under Grant No. 61831002, 61728101, and 61671074, the State Major Science and Technology Special Projects (Grant No. 2018ZX03001023 and 2017ZX03001025-06), and the Beijing Natural Science Foundation under Grant No. JQ18016.

\begin{spacing}{1}

\end{spacing}
\end{document}